\documentclass[twocolumn,preprintnumbers,amsmath,amssymb]{revtex4}

\usepackage{mathrsfs}
\usepackage{amssymb}
\usepackage{graphicx}
\usepackage{dcolumn}
\usepackage{bm}
\usepackage{textcomp}
\usepackage{amsmath}
\usepackage[titletoc]{appendix}
\usepackage{tabularx}
\usepackage{amsthm}
\usepackage{mathrsfs}
\usepackage{lipsum}
\usepackage{longtable}
\usepackage{svg}

\newcommand\ket[1]{\ensuremath{|#1\rangle}}

\newcounter{RomanNumber}
\newcommand{\MyRoman}[1]{\setcounter{RomanNumber}{#1}\Roman{RomanNumber}}

\begin{document}
	\title{ General theory of quantum fingerprinting network}
	\author{Ji-Qian Qin$^1$, Jing-Tao Wang$^1$, Yun-Long Yu$^1$, Xiang-Bin Wang$ ^{1,2,3,4}$}
	
	\affiliation{ \centerline{$^{1}$ State Key Laboratory of Low Dimensional Quantum Physics, Department of Physics,}
	\centerline{Tsinghua University, Beijing 100084, China}
	\centerline{$^{2}$ Shanghai Branch, CAS Center for Excellence and Synergetic Innovation Center in Quantum Information and Quantum Physics,}
	\centerline{ University of Science and Technology of China, Shanghai 201315, China}
	\centerline{$^{3}$ Jinan Institute of Quantum technology, SAICT, Jinan 250101, China}
	\centerline{$^{4}$ Shenzhen Institute for Quantum Science and Engineering, and Physics Department,}
	\centerline{ Southern University of Science and Technology,
	Shenzhen 518055, China}}


	\begin{abstract}
	 The purpose of fingerprinting is to compare long messages with low communication complexity. Compared with its classical version, the quantum fingerprinting can realize exponential reduction in communication complexity. Recently, the multi-party quantum fingerprinting is studied on whether the messages from many parties are the same. However, sometimes it's not enough just to know whether these messages are the same, we usually need to know the relationship among them. We provide a general model of quantum fingerprinting network, defining the relationship function $f^{\rm{R}}$ and giving the corresponding decision rules. In this work, we take the four-party quantum fingerprinting protocol as an example for detailed analysis. We also choose the optimal parameters to minimize communication complexity in the case of asymmetric channels. Furthermore, we compare the multi-party quantum fingerprinting with the protocol based on the two-party quantum fingerprinting and find that the multi-party protocol has obvious advantages, especially in terms of communication time. Finally, the method of encoding more than one bit on each coherent state is used to further improve the performance of the protocol.
	\end{abstract}
	
	\maketitle

    \section{Introduction}

    The application of quantum mechanics in the field of communication brings benefits in many aspects, such as security and communication complexity, which is the minimum amount of communication required among participants in order to complete a task. With respect to security, Quantum key distribution (QKD) \cite{bennett1984quantum,gisin2002quantum,wang2007gaussian,gisin2007quantum,Scarani2009security,Pirandola2019Advances,Xu2020secure} is a representative example that can provide information-theoretic security, while the security of their classical versions is based on computational complexity.  In terms of communication complexity \cite{buhrman2010,buhrman1999multiparty}, compared with the classical version \cite{newman1996public,Babai5678990}, quantum fingerprinting can exponentially reduce the amount of information required to communicate, which is very useful in energy-saving communication \cite{Buhrman2001Quantum,massar2005quantum,Buhrman2010Nonlocality}.

	The purpose of the fingerprinting protocol is to compare the two messages $x_1,x_2\in\{0,1\}^n$, where $n$ is the length of the messages, through the transmission of their fingerprints. In the simultaneous message passing model \cite{yao1979some}, there are two senders that send the fingerprints of the original messages to the Referee, who needs to determine whether $x_1=x_2$ within a small error probability $P_e$. The quantum fingerprinting can significantly reduce the communication complexity required to complete the comparison, i.e., the communication complexity of the classical version is $\mathrm{O}(\sqrt{n})$-bits \cite{newman1996public,Babai5678990}, while that of the quantum version is only $\mathrm{O}(\log_2n)$-qubits \cite{Buhrman2001Quantum}. However, the original protocol requires high-dimensional entanglements, which makes its implementation challenging \cite{Buhrman2001Quantum}. Fortunately, a quantum fingerprinting protocol based on coherent states is proposed \cite{arrazola2014quantum} and since then, there have been many advances in theory and experiment \cite{xu2015experimental,guan2016observation,lovitz2018families}. Recently, in order to reduce the communication time, wavelength division multiplexing (WDM) is used in the quantum fingerprinting and the result shows that not only the communication time is reduced, but also the communication complexity is further reduced \cite{zhong2020efficient}.

    Most studies focus on two-party quantum fingerprinting protocols \cite{arrazola2014quantum,xu2015experimental,guan2016observation,zhong2020efficient} and there are only two relationships between the two inputs, $x_1=x_2$ or $x_1\neq x_2$. As the amount of messages increases, things get more complicated and interesting. If we need to determine whether $N$ messages are the same, a simple way is to use the two-party quantum fingerprinting to compare $N$ inputs in pairs, which needs to be done $(N-1)$ times. So is there a more efficient way? Recently, a multi-party quantum fingerprinting is proposed \cite{GSMultiparty2020}, which only needs to be run once to determine whether $N$ inputs are the same. Obviously, in this case, the quantum fingerprinting network presents a huge advantage. However, the relationship among $N$ inputs is more complex than that between two inputs, and sometimes it is not enough just to know they are different. Usually, we also need to know which of these inputs are different and whether, and by how much, the quantum fingerprinting network still has advantages in this case. Unfortunately, the original protocol \cite{GSMultiparty2020} can not answer these questions. Therefore, it is of great significance to study the quantum fingerprinting network deeply and solve these problems. In this work, based on the existing framework \cite{GSMultiparty2020}, we solve the above problems by the ingenious design of the decision rules. We can determine the relationship among $N$ inputs by exchanging the interference positions of senders so that the comparison results can correspond to a unique relationship, rather than several possible relationships. In addition, we consider the more general case of asymmetric channels and provide the optimal parameter which minimizes the communication complexity. Moreover, by comparing with the protocol based on two-party quantum fingerprinting, it can be found that the quantum fingerprint network has obvious advantages, especially in communication time. In order to further improve the performance of the protocol, we also use a multi-bit encoding scheme \cite{lovitz2018families}, which can further reduce the communication complexity and communication time.

    The rest of this paper is arranged as follows. In Sec.\MyRoman{2}, we present a general $N$-party quantum fingerprinting protocol. Then a four-party quantum fingerprinting network based on the balanced beam-splitter (BS) is used as an example to show how to determine the relationship among multiple inputs in Sec.\MyRoman{3}. Moreover, in Sec.\MyRoman{4}., we consider the more general case of asymmetric channels and optimize the parameters in this case. After this, we discuss the advantages of the quantum fingerprinting network and use the multi-bits encoding method to further improve the performance of the protocol in Sec.\MyRoman{5}. Finally, a brief summary is given in Sec.\MyRoman{6}.

	
   \section{$N$-party Quantum fingerprinting protocol }

	\begin{itemize}
		\item [(1)] There are $N$ senders $S_1,S_2,\dots,S_N$, each of which holds a string of messages $x_k\in\{0,1\}^n, k=1,2,\dots,N,$ to be compared, where $n$ represents the length of the input messages, as shown in Fig.\ref{f3}.
		
		\begin{figure}
			\centering
			\includegraphics[scale=0.3]{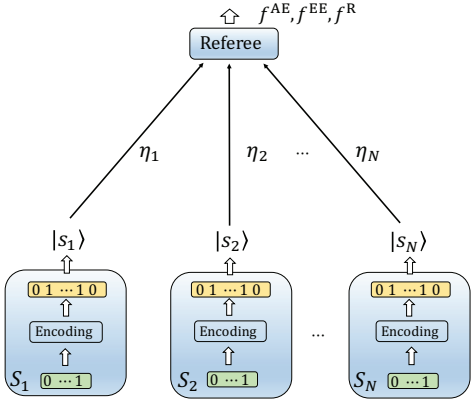}
			\caption{$N$-party quantum fingerprinting protocol. $S_1$, $S_2$, $\dots$, $S_N$ are $N$ senders. They encode the original messages $x_k\in\{0,1\}^n$, which is shown in green, with the ECC and change them into $E(x_k)\in\{0,1\}^m$, which is shown in yellow, $k=1,2,\dots,N$. Then they prepare fingerprint states $\ket{s_k}$ according to $E(x_k)$, and send to the Referee, respectively. $\eta_k$ represents the loss of fingerprint states $\ket{s_k}$ to the Referee. Finally, the Referee can calculate the function $f^{\rm{AE}}$, $f^{\rm{EE}}$, and $f^{\rm{R}}$ according to the different decision rules.}
			\label{f3}
		\end{figure}
		
		\item [(2)] The $N$ senders use the error correction code (ECC) $E (x_k)$ $\in \{0,1\}^m $ to encode input messages, where $m = cn$ is the length of the $E(x_k)$, $c>1$, and $\delta$ is the smallest relative Hamming distance of any two different messages. The ECC is used to increase the hamming distance between different inputs. For $k\neq k^{\prime}$, $E(x_k)$ and $E(x_{k^{\prime}})$ have at least $\delta m$ different bits.

		\item [(3)] Each sender $S_k$ encodes the coherent state according to $E(x_k)$, then sends them to the Referee. 
		
		\begin{equation}\label{e1}
		\begin{split}
		&\ket{s_k}=\bigotimes_{j=1}^{m}\biggl |(-1)^{E(x_k)_j}\frac{\alpha_k}{\sqrt{m}}\biggr >_j\\
		\end{split}
		\end{equation}
		We can call $\ket{s_k}$ the fingerprint state, which contains $m$ coherent states. 
		The amplitude of each coherent state is $\frac{\alpha_k}{\sqrt{m}}$ and the phase depends on $E(x_k)_j$.

		In practice, the impact of the channel loss on the protocol needs to be considered, so we use the $\eta_k$ to represent the losses experienced by these fingerprint states respectively. If the channel is symmetric, $\eta_k=\eta$. More generally, the loss of different channels is not equal, which will affect the selection of optimal parameters.
		
		\item [(4)] The Referee uses the device consisting of several BSs and $N$ detectors $\rm{D_1,D_2,\dots,D_N}$ to perform the interference measurement with the $N$ fingerprint states. 
		
		\item [(5)] The Referee records the total counts on $N$ detectors, which are $C_1, C_2,\dots,C_N$, and selects appropriate thresholds $C_1^{\rm{th}}, C_2^{\rm{th}},\dots,C_N^{\rm{th}}$. Then, the Referee compares $C_k$ with $C_k^{\rm{th}}$. If $C_k<C_k^{\rm{th}}$, the Referee records the result as the number $0$, otherwise, records it as the number $1$.
		
		\item [(6)] The Referee can draw a conclusion about the following contents under the condition that error probability $P_e\le \varepsilon$ ($\varepsilon$ is the maximum error probability that the protocol can tolerate) according to the corresponding decision rules.
		
	\end{itemize}

		\textbf{All-equality function} \cite{fischer2016Public} : Determine whether the $N$ messages are the same,
		
		\begin{equation}\label{fae}
		f^{\rm{AE}}=\left\{
		\begin{array}{lr}
		1, & x_1=x_2=\dots=x_N\\
		0, &  \rm{else}.\\
		\end{array}
		\right.
		\end{equation}
		
		\textbf{Exists-equality function} \cite{fischer2016Public}: Determine whether there is an equal pair of input messages,
		
		\begin{equation}\label{fee}
		f^{\rm{EE}}=\left\{
		\begin{array}{lr}
		1, &  \mathrm{for\ some}\ k\neq{k}^{\prime},\ x_k=x_{k'}\\
		0, &  \rm{else}.\\
		\end{array}
		\right.
		\end{equation}
				
		\textbf{Relationship function} : Determine the relationship among the $N$ input messages,
		
		\begin{equation}\label{fR}
		\begin{split}
		&f^{\rm{R}}(x_1,x_2,\dots,x_N)=\\
		&\left\{
		\begin{array}{lr}
		v^{N,1}_1, & (x_1,x_2,\dots,x_N)_1\\
		v^{N-1,2}_1, & (x_1,\dots,x_{N-1})_1,(x_N)_2\\
		\vdots & \vdots\\
		v^{N-1,2}_{C_N^1}, & (x_2,\dots,x_{N})_1,(x_1)_2\\
		v^{N-2,2}_1, & (x_1,\dots,x_{N-2})_1,(x_{N-1},x_{N})_2 \\
		\vdots & \vdots\\
		v^{N-2,2}_{C_N^2}, & (x_3,\dots,x_{N})_1,(x_{1},x_{2})_2 \\
		\vdots & \vdots\\
		v^{i,j}_{k}, & (G_1)_1,(G_2)_2,\dots,(G_j)_j  \\
		\vdots & \vdots\\
		v^{1,N}_1, & (x_1)_1,(x_2)_2,\dots,(x_N)_N,\\
		\end{array}
		\right.
		\end{split}
		\end{equation}
		where $v^{i,j}_k$ represents different values of $f^{\rm{R}}$ and we write different inputs in different parentheses, where subscripts are used to distinguish the different parentheses. There are $j$ such groups and $G_j$ represents the same $N_j$ inputs of the total $N$ inputs, which satisfies $N_1\ge N_2\ge\dots\ge N_j$ and $\sum_{j_0=1}^{j} N_{j_0} =N$. We define $N_1=i$, which means at most $i$ inputs are the same, $i=N-(j-1),N-(j-1)-1,\dots,\lceil \frac{N}{j} \rceil$. $k=1,2,\dots,k_{t}^{i,j}$, where $k_{t}^{i,j}$ is the total number of cases for the same $(i,j)$. We provide the expression for $k_{t}^{i,j}$ in the Appendix \ref{A}.

		In function $f^{\rm{R}}$, for example, $v_1^{N-1,2}$ represents the first relationship among $N$ inputs when there are two groups and at most $(N-1)$ inputs are the same, i.e., $(x_1,\dots,x_{N-1})_1,(x_N)_2$. When $N=4$, the form of $f^{\rm{R}}$ is shown in the last column of Table \ref{t9}.

		\textbf{Decision rules of the $f^{\rm{AE}}:$}
		The device in the Referee has the feature that there is only one detector $\rm{D_1}$ has clicks if all inputs are equal and all detectors are likely to have clicks as long as there are different inputs, ideally. The Referee can determine whether the $N$ inputs are equal, according to the results of comparison between total counts on detectors and the threshold $C_{1}^{\rm{th}}$ or $C_{\rm{sum}}^{\rm{th}}$ \cite{GSMultiparty2020}. If $C_1>C_1^{\rm{th}}$ or $C_{\rm{sum}}<C_{\rm{sum}}^{\rm{th}}$, where $C_{\rm{sum}}=C_2+\dots+C_{N}$, $N$ inputs are equal ($f^{\rm{AE}}=1$). If $C_1<C_1^{\rm{th}}$ or $C_{\rm{sum}}>C_{\rm{sum}}^{\rm{th}}$, $N$ inputs are different ($f^{\rm{AE}}=0$).

        \textbf{Decision rules of the $f^{\rm{R}}:$} 
        The Referee can determine the relationship among $N$ inputs based on the comparison results of $C_k$ and $C_k^{\rm{th}}$. If the result of the first run of the protocol corresponds to a unique relationship, then the calculation of $f^R$ is complete. Otherwise, we need to exchange the interference positions of senders and run the protocol several times until the comparison results can correspond to a unique relationship, rather than several possible relationships. As shown in Table \ref{t9}, in this work, we analyze in detail how to calculate $f^{\rm{R}}$ in the four-party quantum fingerprinting protocol ($N=4$). 
        
        From Eq.\eqref{fae}-Eq.\eqref{fR}, we can find that as long as we calculate $f^{\rm{R}}$, we can know the values of $f^{\rm{AE}}$ and $f^{\rm{EE}}$, and conversely, this is not the case, so the calculation of $f^{\rm{R}}$ is very important. In the following, we mainly consider the calculation of $f^{\rm{R}}$.
        
\begin{figure}
	\centering
	\includegraphics[scale=0.32]{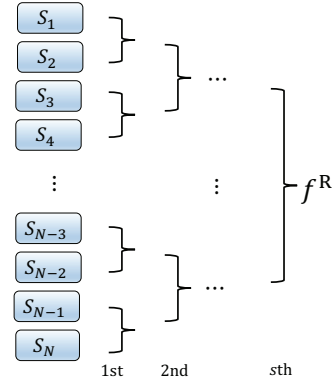}
	\caption{Determine the relationships among $N$ messages based on the quantum fingerprinting network composed of balanced BSs, $N=2^s, s\in \mathbb{Z^+}$. $S_1$, $S_2$,\dots,$S_N$ are $N$ senders who hold the messages to be compared. Each bracket, which is equivalent to a balanced BS in the Fig. \ref{f1}, compares the input of the two parts. The first layer shows the comparison results of $E(x_k)$ and $E(x_{k+1})$, $k=1,3,\dots,N-1$; the second layer shows the comparison results of $E(x_k)+E(x_{k+1})$ and $E(x_{k+2})+E(x_{k+3})$, $k=1,5,\dots,N-3$, and so on, and the $s$th layer shows the comparison results of $\sum_{k=1}^{2^{s-1}}E(x_k)$ and $\sum_{k=2^{s-1}+1}^{N}E(x_k)$. }
	\label{f4}
\end{figure}

\textbf{The design of $N$-party quantum fingerprinting network based on balanced BSs.}
We show in Fig.\ref{f4} how to determine the relationships among $N$ inputs based on quantum fingerprinting network composed of balanced BSs, $N=2^s, s\in \mathbb{Z^+}$. Each bracket, which is equivalent to the BS in the Fig.\ref{f1}, compares the input of the two parts. The first layer shows the comparison results of $E(x_i)$ and $E(x_{i+1})$, $i=1,3,\dots,N-1$; the second layer shows the comparison results of $E(x_i)+E(x_{i+1})$ and $E(x_{i+2})+E(x_{i +3})$, $i=1,5,\dots,N-3$, and so on, and the $s$th layer shows the comparison results of $\sum_{i=1}^{2^{s-1}}E(x_i)$ and $\sum_{i=2^{s-1}+1}^{N}E(x_i)$. Swapping interference positions is equivalent to changing the content of the comparison to get a new set of comparison relationships. We need to swap the interference position of different senders until we can determine a unique relationship, not just a few possible relationships, based on the comparison results. 

\textbf{Communication Complexity.}
When determining whether the $N$ inputs are equal (calculating $f^{\rm{AE}}$), the total communication complexity of the classical optimal fingerprinting protocol $C_{\rm{o}}^{\rm{AE}}$ and the classical limit $C_{\rm{l}}^{\rm{AE}}$ are as follows \cite{fischer2016Public,GSMultiparty2020}

\begin{equation}\label{}
\begin{split}
&C_{\mathrm{o}}^{\rm{AE}}=\\
&N\lceil\frac{\log_2(P_e)}{\log_2\left[1-\frac{1}{9}(1-e^{-\frac{1}{2}})\right]}\rceil\times \left[8\sqrt{2\lceil \frac{3n}{N} \rceil}+4\lceil \log_2(3n/\lceil\frac{3n}{N}\rceil)\rceil\right],\\
\end{split}
\end{equation}

\begin{equation}\label{}
\begin{split}
C_{\rm{l}}^{\rm{AE}}&=N\left[\frac{(1-2\sqrt{P_e})\sqrt{n}}{2\sqrt{N \ln 2}}-\frac{1}{N}\right],\\
\end{split}
\end{equation}
where $P_e$ is the error probability of the fingerprinting protocol.  

Determining the relationship among multiple inputs is more difficult than determining whether they are equal, so the following relationship should exist

\begin{equation}\label{QC}
\begin{split}
&C_{\rm{o}}^{\rm{R}}>C_{\rm{o}}^{\rm{AE}},\\ &C_{\rm{l}}^{\rm{R}}>C_{\rm{l}}^{\rm{AE}},\\
\end{split}
\end{equation}
where $C_{\rm{o}}^{\rm{R}}$ and $C_{\rm{l}}^{\rm{R}}$ represent the optimal communication complexity in classical version and classical limit when calculating $f^{\rm{R}}$, respectively. 

In quantum fingerprinting based on coherent states \cite{arrazola2014quantum}, the total communication complexity $Q^{\rm{AE}}$ when calculating $f^{\rm{AE}}$ can be written as 

\begin{equation}\label{}
\begin{split}
Q^{\rm{AE}}&=\sum_{k=1}^N\mu_{k}\log_2 n ,\\
\end{split}
\end{equation}
where $\mu_k=|\alpha_k|^2$ represents the total photon number of the fingerprint state.
According to the decision rules of $f^{\rm{R}}$, the maximum total  communication complexity $Q^{\rm{R}}$ can be written as

\begin{equation}\label{}
\begin{split}
Q^{\rm{R}}&=\sum_{t_0=1}^{t_{\rm{max}}} Q^{\rm{AE}}_{t_0} ,\\
\end{split}
\end{equation}
where $t_{\rm{max}}$ represents the maximum number of times the protocol need to be run in order to calculate $f^{\rm{R}}$.

In order to compare the performance of the multi-party quantum fingerprinting protocol with the classical version in terms of communication complexity when calculating $f^{\rm{R}}$, we need to compare $C_{\rm{o}}^{\rm{R}}$, $C_{\rm{l}}^{\rm{R}}$ and $Q^{\rm{R}}$. However, there are very few studies on the communication complexity $C_{\rm{o}}^{\rm{R}}$ and $C_{\rm{l}}^{\rm{R}}$. Therefore, in this work, we choose to compare $C_{\rm{o}}^{\rm{AE}}$, $C_{\rm{l}}^{\rm{AE}}$ and $Q^{\rm{R}}$. When $Q^{\rm{R}}<C_{\rm{l}}^{\rm{AE}}<C_{\rm{o}}^{\rm{AE}}$, then combined with \eqref{QC}, it is obvious that $Q^{\rm{R}}<C_{\rm{l}}^{\rm{R}}<C_{\rm{o}}^{\rm{R}}$ can be concluded, which can reflect the advantage of multi-party quantum fingerprint protocol in communication complexity.


	\section{ Determine the relationship in Four-party Quantum Fingerprinting }

\begin{figure}
    	\centering
    	\includegraphics[scale=0.25]{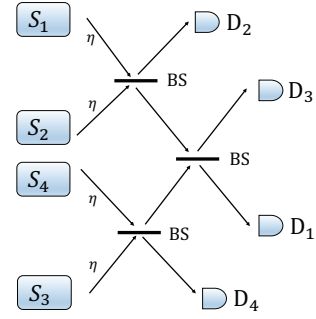}
    	\caption{Four-party quantum fingerprinting of symmetric channels ($\eta_k=\eta, k=1,2,3,4$) based on balanced BSs. $S_1$, $S_2$, $S_3$, $S_4$ are four senders, who hold the messages to be compared. $\rm{D_1}$, $\rm{D_2}$, $\rm{D_3}$ and $\rm{D_4}$ are four detectors in the Referee, who can determine the relationship among four inputs according to the decision rules of $f^{\rm{R}}$.}
    	\label{f1}
    \end{figure}

    Consider the case of four-party quantum fingerprinting protocol with symmetric channel, where $\eta_k=\eta$ and $\mu_k=|\alpha_k|^2=\mu$ , $k=1,2,3,4$, as shown in Fig. \ref{f1}. 
    
\begin{figure}
    	\centering
    	\includegraphics[scale=0.28]{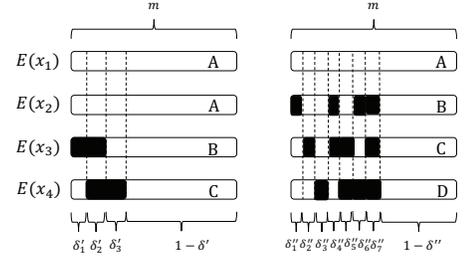}
    	\caption{In the four-party quantum fingerprinting protocol, the left relationship is AABC and the right relationship is ABCD. $m$ is the length of the $E(x_k), k=1,2,3,4$, and $\delta^\prime$ and $\delta^{\prime \prime}$ represent the relative Hamming distance. In the relationship of AABC,  $\delta^{\prime}=\sum_{i=1}^{3}\delta_i^{\prime}$. In the relationship of ABCD, $\delta^{\prime\prime}=\sum_{i=1}^{7}\delta_i^{\prime\prime}$.}
    	\label{f2}
    \end{figure}

    There are $15$ kinds of relationships among the four inputs $x_1$, $x_2$, $x_3$ and $x_4$, as shown in the first column of Table \ref{t9}. The relationship among $\{x_k\}$ is the same as the relationship among $\{E(x_k)\}$. The same letters represent the same inputs and different letters represent different inputs. For example, $\rm{AAAA}$ means that the four inputs are equal and $\rm{ABCD}$ means that they are completely different. Among them, AABC and ABCD are more complicated, as analyzed in Fig. \ref{f2}. $E(x_1)$ is all white, indicating the reference message. The white part of $E(x_2),E(x_3)$ and $E(x_4)$ represents the same part to $E(x_1)$, and the black part represents the different part to $E(x_1)$. For the convenience of expression, $E(x_k)$ are shown as above. The order of the actual $E(x_k)$ may not be the same as the Fig. \ref{f2}, but we only care about the total counts $C_{k}$ on the detectors, so the order does not affect the decision rules. In $\rm{AABC}$, $\delta^{\prime}=\delta_1'+\delta_2'+\delta_3'$. Since $\delta$ is the  minimum relative distance of ECC, it satisfies  $\min\{\delta_1'+\delta_2', \delta_1'+\delta_3',  \delta_2'+\delta_3'\}\ge \delta$, which means that the distance between $E(x_k)$ and $E(x_{k'})$ is at least $\delta m$ when $k\neq k'$. In $\rm{ABCD}$,
    $\delta^{''}=\sum_{i=1}^{7}\delta_i''$. In the same way, $\min\{\delta_1''+\delta_4''+\delta_6''+\delta_7'',\delta_2''+\delta_4''+\delta_5''+\delta_7'',\delta_3''+\delta_5''+\delta_6''+\delta_7'',\delta_1''+\delta_2''+\delta_5''+\delta_6'',\delta_1''+\delta_3''+\delta_4''+\delta_5'', \delta_2''+\delta_3''+\delta_4''+\delta_6''\}\ge \delta.$

    \begin{table}
    	\centering
    	\caption{Determine the relationship among the four messages $x_1,x_2,x_3,x_4$ by three detectors $\rm{D_2}$, $\rm{D_3}$ and $\rm{D_4}$ in Fig.\ref{f1}. $\rm{R1}$ represents the comparison results of total counts $C_2,C_3$, $C_4$ and thresholds $C_{2}^{\rm{th}},C_{3}^{\rm{th}}$, $C_{4}^{\rm{th}}$ of the three detectors after the first run of the protocol. For example, if $C_2<C_2^{\rm{th}}$, $C_3<C_3^{\rm{th}}$, $C_4<C_4^{\rm{th}}$, we record it as R1$=000$. Similarly, $\rm{R2}$ and $\rm{R3}$ represent the comparison results of the three detectors after the second and third run of the protocol after exchanging the interference position of $S_2$,$S_3$ and $S_3$,$S_4$, respectively. $\widetilde{\rm{R1}}$, $\widetilde{\rm{R2}}$, $\widetilde{\rm{R3}}$,
        represent the comparison results corresponding to the relationship of ABCD in the three runs. [$\widetilde{\rm{R1}}$, $\widetilde{\rm{R2}}$, $\widetilde{\rm{R3}}$]$\in \{[101,111,\emptyset], [111,101,\emptyset], [111,111,101], [111,111,111]\}$, where $\emptyset$ means that there is no need for a third comparison. In $f^{\rm{R}}$, we write different inputs in different parentheses.}
    	\begin{tabular}{p{1.3cm}p{0.9cm}p{0.9cm}p{0.9cm}p{3.8cm}}
    		\hline
    		& R1 & R2 &  R3   & $f^{\rm{R}}$ \\
    		\hline
    		\hline                  
    		AAAA & $000$ & &  & $14:(x_1,x_2,x_3,x_4)_1$\\
    		AAAB & $011$ &$011$ && $ 13:(x_1,x_2,x_3)_1,(x_4)_2$\\
    		AABA & $011$ & $110$& & $ 12:(x_1,x_2,x_4)_1,(x_3)_2$\\	
    		ABAA & $110$ &$011$&& $ 11:(x_1,x_3,x_4)_1,(x_2)_2$\\
    		BAAA & $110$ &$110$ & & $ 10:(x_2,x_3,x_4)_1,(x_1)_2$\\		
    		AABB & $010$ & && $9:(x_1,x_2)_1,(x_3,x_4)_2$\\
    		ABAB & $101$ & $010 $ & &$ 8:(x_1,x_3)_1,(x_2,x_4)_2$\\
    		ABBA & $101$ & $101$&&$7:(x_1,x_4)_1,(x_2,x_3)_2$\\
    		AABC & $011$ &$111$ & &$ 6:(x_1,x_2)_1,(x_3)_2,(x_4)_3$\\
    		ABAC & $111$ & $011$ & &$ 5:(x_1,x_3)_1,(x_2)_2,(x_4)_3$\\
    		ABCA & $111$ & $111$ & $011$ &$4:(x_1,x_4)_1,(x_2)_2,(x_3)_3$\\
    		BAAC & $111$ & $111$ & $110$ &$3:(x_2,x_3)_1,(x_1)_2,(x_4)_3$\\
    		BACA & $111$ & $110$ & &$ 2:(x_2,x_4)_1,(x_1)_2,(x_3)_3$\\
    		BCAA & $110$ & $111$ & &$ 1:(x_3,x_4)_1,(x_1)_2,(x_2)_3$\\		
    		ABCD & $\widetilde{\rm{R1}}$ & $\widetilde{\rm{R2}}$ & $\widetilde{\rm{R3}}$&$ 0:(x_1)_1,(x_2)_2,(x_3)_3,(x_4)_4$\\
    		\hline          
    	\end{tabular}\label{t9}
    \end{table}
    
    Based on the above analysis, we can list the probability of observing a click on detectors $\rm{D_1}$, $\rm{D_2}$, $\rm{D_3}$, and $\rm{D_4}$ for each pulse sent under all relationships, as shown in table \ref{t1}. $\delta_{\rm{tol}}$ represents the proportion of events except for $E(x_1)_j=E(x_2)_j=E(x_3)_j=E(x_4)_j$, $j\in[1,m]$, in $E(x_k), k=1,2,3,4$. $\delta_{1,2}$ represents the relative distance between $E(x_1)$ and $E(x_2)$. Similarly, $\delta_{3,4}$ represents the relative distance between $E(x_3)$ and $E(x_4)$. $\delta_{12,34}^1$ represents the proportion of events, where the relationship among $E(x_1)_j, E(x_2)_j, E(x_3)_j, E(x_4)_j\in$  $\{\mathrm{AAAB, AABA, ABAA, BAAA}\}$, in entire $E(x_k)$; $\delta_{12,34}^2$ represents the proportion of events, where the relationship among $E(x_1)_j, E(x_2)_j, E(x_3)_j, E(x_4)_j\in\{\rm{AABB}\}$, in entire $E(x_k)$. For example, if the relationship is AABC,
    
    \begin{equation}
    \begin{split}
    &P_{\rm{D_1}}=(1-\delta')(1-e^{-\frac{4\eta\mu}{m}})+(\delta_1'+\delta_3')(1-e^{-\frac{\eta\mu}{m}})+P_d,\\ &P_{\rm{D_2}}=P_d, \\
    &P_{\rm{D_3}}=(\delta_1'+\delta_3')(1-e^{-\frac{\eta\mu}{m}})+\delta_2'(1-e^{-\frac{4\eta\mu}{m}})+P_d,\\ &P_{\rm{D_4}}=(\delta_1'+\delta_3')(1-e^{-\frac{2\eta\mu}{m}})+P_d,
    \end{split}
    \end{equation} 
    where $P_d$ represents the dark count probability of detectors.
    
    We can calculate $f^{\rm{R}}$ according to the probability of observing a click on detectors $\rm{D_2}$, $\rm{D_3}$ and $\rm{D_4}$ in Fig.\ref{f1}.

    \begin{table*}  
    	\vspace{20pt}           
    	\centering
    	\caption{The probability $P_{\rm{D_1}}$, $P_{\rm{D_2}}$, $P_{\rm{D_3}}$,$P_{\rm{D_4}}$ of observing a click on detectors $\rm{D_1, D_2, D_3, D_4}$ under all relationships in four-party quantum fingerprinting. $\delta_{\rm{tol}}$ represents the proportion of events except for $E(x_1)_j=E(x_2)_j=E(x_3)_j=E(x_4)_j$, $j\in[1,m]$, in $E(x_k), k=1,2,3,4$. $\delta_{1,2}$ represents the relative distance between $E(x_1)$ and $E(x_2)$. Similarly, $\delta_{3,4}$ represents the relative distance between $E(x_3)$ and $E(x_4)$. $\delta_{12,34}^1$ represents the proportion of events, where the relationship among $E(x_1)_j, E(x_2)_j, E(x_3)_j, E(x_4)_j$ $\in\{\rm{AAAB, AABA, ABAA,BAAA}\}$ in entire $E(x_k)$; $\delta_{12,34}^2$ represents the proportion of events, where the relationship among $E(x_1)_j, E(x_2)_j, E(x_3)_j, E(x_4)_j\in$ $\{\mathrm{AABB}\}$, in entire $E(x_k)$. }
    	\begin{tabular}{p{6cm}|p{3cm}|p{5.5cm}|p{2.9cm}}
    		\hline
    		 $ P_{\rm{D_1}} $ & $ P_{\rm{D_2}}$ &  $ P_{\rm{D_3}} $  &  $ P_{\rm{D_4}}$  \\
    		\hline 
    		\hline                 
    	    $(1-\delta_{\rm{tol}})(1-e^{-\frac{4\eta\mu}{m}})+\delta^1_{12,34}(1-e^{-\frac{\eta\mu}{m}})+P_d$ 
    	    & $ \delta_{1,2}(1-e^{-\frac{2\eta\mu}{m}})+P_d$
    	    & $ \delta_{12,34}^1(1-e^{-\frac{\eta\mu}{m}})+\delta_{12,34}^2(1-e^{-\frac{4\eta\mu}{m}})+P_d $  & $ \delta_{3,4}(1-e^{-\frac{2\eta\mu}{m}})+P_d$ \\
    		\hline
    	\end{tabular}\label{t1}
    \end{table*}

    \textbf{Decision rules of $f^{R}$ in four-party quantum fingerprinting}: 
    
    Select the threshold $C_{2}^{\rm{th}}$, $C_{3}^{\rm{th}}$, $C_{4}^{\rm{th}}$ and compare the total counts $C_2$, $C_3$, $C_4$ of the three detectors  with corresponding thresholds respectively. As shown in table \ref{t9}, $\rm{R1}$ represents the comparison results after the first run of the protocol. The number $0$ means the total counts are less than the threshold, and the number $1$ means the total counts are greater than the threshold. For example, if $C_2<C_2^{\rm{th}}$, $C_3<C_3^{\rm{th}}$ and $C_4<C_4^{\rm{th}}$, we record it as R1=000. On the first run of the protocol, if the result on $\rm{D_2}$ is $0$, then it means $E(x_1)=E(x_2)$, otherwise, it means $E(x_1)\neq E(x_2)$; similarly, if the result on $\rm{D_4}$ is $0$, then it means $E(x_3)=E(x_4)$, otherwise, it means $E(x_3)\neq E(x_4)$; if the result on $\rm{D_3}$ is $0$, then it means $E(x_1)+E(x_2)=E(x_3)+E(x_4)$, otherwise, it means $E(x_1)+E(x_2)\neq E(x_3)+E(x_4)$. When we run the protocol once, we can identify two of the relationships, i.e., $f^{\rm{R}}=14$ (AAAA) and $f^{\rm{R}}=9$ (AABB), for which the results of comparison are R1=$000$ and R1=$010$, respectively. If R1=011, then it corresponds to three possible relationships AAAB, AABA and AABC. To determine which relationship it is, we need to swap the interference positions of $S_2$ and $S_3$ and run the protocol for the second time.
    
    On the second run of the protocol, if the result on $\rm{D_2}$ is $0$, then it means $E(x_1)=E(x_3)$, otherwise, it means $E(x_1)\neq E(x_3)$; similarly, if the result on $\rm{D_4}$ is $0$, then it means $E(x_2)=E(x_4)$, otherwise, it means $E(x_2)\neq E(x_4)$; if the result on $\rm{D_3}$ is $0$, then it means $E(x_1)+E(x_3)=E(x_2)+E(x_4)$, otherwise, it means $E(x_1)+E(x_3)\neq E(x_2)+E(x_4)$. For instance, if R1=011, R2=011, then $f^{\rm{R}}=13$ (AAAB). If we still can not determine the unique relationship according to R1 and R2 (for example, R1=R2=111), we need to exchange the interference positions of $S_3$ and $S_4$ on the basis of the second swapping, and then run the protocol for the third time. 
    
    On the third run of the protocol, if the result on $\rm{D_2}$ is $0$, then it means $E(x_1)=E(x_4)$, otherwise, it means $E(x_1)\neq E(x_4)$; similarly, if the result on $\rm{D_4}$ is $0$, then it means $E(x_2)=E(x_3)$, otherwise, it means $E(x_2)\neq E(x_3)$; if the result on $\rm{D_3}$ is $0$, then it means $E(x_1)+E(x_4)=E(x_2)+E(x_3)$, otherwise, it means $E(x_1)+E(x_4)\neq E(x_2)+E(x_3)$. If R1=111, R2=111, R3=011, then $f^{\rm{R}}=4$ (ABCA).  
    $\widetilde{\rm{R1}}$, $\widetilde{\rm{R2}}$, $\widetilde{\rm{R3}}$
    respectively represent the comparison results corresponding to ABCD in the three runs. 
    [$\widetilde{\rm{R1}}, \widetilde{\rm{R2}}, \widetilde{\rm{R3}}$]$\in$$\{[101,111,\emptyset]$, $[111,101,\emptyset]$, $[111,111,101]$, $[111,111,111]\}$, where $\emptyset$ means that there is no need for the third comparison. In other words, if the relationship among the four inputs is ABCD, it can be identified by running the protocol twice or three times and whether a third run is needed depends on R1 and R2.
    
    In short, we can determine the relationship among four inputs by exchanging the interference positions of senders to compare different parts. Whether the latter run is needed depends on the previous results. That is to say, whether the second run is needed depends on R1, and whether the third run is needed depends on R1 and R2. When the comparison result corresponds to only one relationship, instead of several possible relationships, the calculation of $f^{\rm{R}}$ is completed.

    Then, we show how to choose appropriate thresholds $C_{2}^{\rm{th}}$, $C_{3}^{\rm{th}}$ and $C_{4}^{\rm{th}}$ and total photon number $\mu$ of the fingerprint state. The total number of clicks on the detectors is approximated by a binomial distribution $C_{k,u}\sim\rm{Bin}(\emph{m},\emph{P}_\emph{k}^\emph{u})$, where $k\in\{2,3,4\}$ and $u\in \{\mathrm{E,D}\}$.
    The probabilities of observing a click on three detectors $\rm{D_2}$, $\rm{D_3}$ and $\rm{D_4}$ are as follow when the inputs are equal and different, 
    
    \begin{equation}\label{}
    \begin{split}
    P_{2}^{\rm{E}}&=P_d\\
    P_{2}^{\rm{D}}&=\delta(1-e^{-\frac{2\eta\mu}{m}})+P_d\\
    P_{3}^{\rm{E}}&=P_d\\
    P_{3}^{\rm{D}}&=\delta(1-e^{-\frac{\eta\mu}{m}})+P_d\\
    P_{4}^{\rm{E}}&=P_d\\
    P_{4}^{\rm{D}}&=\delta(1-e^{-\frac{2\eta\mu}{m}})+P_d,\\
    \end{split}
    \end{equation}
    where $P_d$ represents the dark count probability of detectors, $\mu$ is the total mean photon number of the fingerprint state, $\delta$ represents the minimum relative distance of ECC and $\eta$ describes the channel loss.
    
    The error probability $P_{e}$ of the protocol can be defined as follows
    
    \begin{equation}\label{}
    \begin{split}
    P_{e}=\max\{&P^{\rm{E}}(C_{2}>C_{2}^{\rm{th}}), P^{\rm{D}}(C_{2}<C_{2}^{\rm{th}}), \\
    &P^{\rm{E}}(C_{3}>C_{3}^{\rm{th}}), P^{\rm{D}}(C_{3}<C_{3}^{\rm{th}}),\\
    &P^{\rm{E}}(C_{4}>C_{4}^{\rm{th}}), P^{\rm{D}}(C_{4}<C_{4}^{\rm{th}})\}.\\
    \end{split}
    \end{equation}
   
   With the proper selection of  $\mu$ and threshold $C_2^{\rm{th}}$, $C_3^{\rm{th}}$ and $C_4^{\rm{th}}$, we can make $Q^{\rm{R}}$ as small as possible under the condition that $P_e\le \varepsilon$.
   If $\mu$ is too small, $Q^{\rm{R}}$ can be small, but no matter how to choose the threshold, $P_e$ may be very large, which cannot meet the requirements of the protocol. Conversely, if $\mu$ is too large, then $Q^{\rm{R}}$ will be large even if the appropriate threshold can make $P_e$ small. Therefore, we should choose appropriate $\mu$  and $C_{2}^{\rm{th}}$, $C_{3}^{\rm{th}}$ and $C_{4}^{\rm{th}}$ to balance the relationship between $Q^{\rm{R}}$ and $P_e$.

\begin{table}
	\centering
	\caption{ Determine the relationship among four inputs based on three detectors $\rm{D_2}$, $\rm{D_3}$ and $\rm{D_4}$ in Fig.\ref{f1}. $n=10^{13}$, $\delta=0.22$, $\eta=0.1$, $P_d=10^{-11}$, $c=0.2$, $\varepsilon=10^{-2}$. $ C_{\rm{o}}^{\rm{AE}}=1.29\times 10^{10} $, $ C^{\rm{AE}}_{\rm{l}}=3.04\times 10^{6}. $}
	\begin{tabular}{p{1.4cm}p{1.4cm}p{1.4cm}p{1.4cm}p{1.4cm}}
		\hline 
		$ \mu $&$ C_{2}^{\rm{th}} $ &  $ C_{3}^{\rm{th}} $  &  $ C_{4}^{\rm{th}} $   &  $ Q^{\rm{R}} $  \\
		\hline 
		\hline  
		$ 4961 $ &$ 602 $ & $ 553 $ & $602 $  &  $2.57\times 10^6$\\              
		\hline
	\end{tabular}\label{t4}
\end{table}

We can see from table \ref{t4} that when $\eta=0.1$ and $P_e\le10^{-2}$, the total number of photons sent by each sender $\mu=4961$, the thresholds of the three detectors are chosen as $C_2^{\rm{th}}=602$, $C_3^{\rm{th}}=553$ and $C_4^{\rm{th}}=602$, respectively, and the maximum total communication complexity $Q^{\rm{R}}=2.57\times 10^6$, which is four orders of magnitude smaller than the classical optimal protocol $C_{\rm{o}}^{\rm{AE}}=1.29\times 10^{10}$. Meanwhile, $Q^{\rm{R}}$ also breaks the classical limit $C_{\rm{l}}^{\rm{AE}}=3.04\times 10^6$. To sum up, they satisfy $Q^{\rm{R}}<C_{\rm{l}}^{\rm{AE}}<C_{\rm{o}}^{\rm{AE}}$, and combined with \eqref{QC}, we can conclude that $Q^{\rm{R}}<C_{\rm{l}}^{\rm{R}}<C_{\rm{o}}^{\rm{R}}$. This shows the advantages of the quantum fingerprint network in communication complexity compared with the classical version.

In conclusion, as shown in Table \ref{t9}, we provide an efficient way to determine the relationships among four messages and we can easily obtain the values of $f^{\rm{AE}}$ and $f^{\rm{EE}}$ with the knowledge of $f^{\rm{R}}$.


\section{Quantum fingerprinting protocol of asymmetric channels }

    We consider the more general cases of asymmetric channels. Whether the channel is symmetric does not change the decision rules of the Referee. However, it affects the probability of observing a click on the detectors for each pulse when the inputs are equal and different, which has an impact on choosing the appropriate fingerprint states and thresholds to minimize communication complexity.

	\subsection{Two-party quantum fingerprinting protocol of asymmetric channels}
	
	When $N=2$, $f^{\rm{AE}}=f^{\rm{EE}}=f^{\rm{R}}$, $Q^{\rm{AE}}=Q^{\rm{EE}}=Q^{\rm{R}}$.
	The communication complexity $Q^{\rm{R}}$ should be taken as the objective function to optimize the parameters $\alpha_1$, $\alpha_2$ and $C_2^{\rm{th}}$ under the condition of $P_e\le \varepsilon$.

    \begin{table}
    	\centering
    	\caption{The selection of parameters in two-party quantum fingerprinting protocol of asymmetric channels. $n=3\times10^{12}$, $c=0.2$, $P_d=10^{-10}$, $\delta=0.22$, $\sqrt{\eta_1}=0.3$, $\sqrt{\eta_2}=0.4$, $\varepsilon=10^{-5}$. $C_{\rm{o}}^{\rm{AE}}=1.24\times 10^{10}$, $C_{\rm{l}}^{\rm{AE}}=1.46\times 10^{6}$ .}
    	\begin{tabular}{p{1.5cm}p{1.5cm}p{1.5cm}p{2cm}}
    		\hline 
    		 $\alpha_1$ & $ \alpha_2 $ & $ C_{2}^{\rm{th}} $ & $ Q^{\rm{R}} $ \\
    		\hline 
    		\hline                 
    		$85$& $78$ & $1685$ & $ 5.52\times 10^5$ \\
    		\hline
    	\end{tabular}\label{t15}
    \end{table}

    In Table \ref{t15}, we show the results of parameter optimization in two-party quantum fingerprinting of asymmetric channels. The details are in the Appendix \ref{C}. When $\sqrt{\eta_1}=0.3$ and $\sqrt{\eta_1}=0.4$, the total amplitude of the two fingerprint states can be set to $\alpha_1=85$ and $\alpha_2=78$ respectively, and the threshold value $D_{2}^{\rm{th}}=1685$. At this time, the communication complexity $Q^{\rm{R}}=5.52\times 10^5$. Not only is it about five orders of magnitude less than the classical communication complexity $C_o^{\rm{AE}}=1.24\times 10^{10}$, but it also breaks the classical limit $C_l^{\rm{AE}}=1.46\times 10^6$. They satisfy $Q^{\rm{R}}<C_{\rm{l}}^{\rm{AE}}<C_{\rm{o}}^{\rm{AE}}$ and combined with \eqref{QC}, we can obtain $Q^{\rm{R}}<C_{\rm{l}}^{\rm{R}}<C_{\rm{o}}^{\rm{R}}$. Therefore, compared with the classical version, the two-party quantum fingerprinting has great advantages in terms of communication complexity.

	\subsection{Four-party Quantum fingerprinting of asymmetric channels }

	We show how the Referee can determine the relationship among four inputs of asymmetric channels. The decision rules of the Referee are the same as those of the symmetric channels, as shown in Table \ref{t9}. The difference lies in the probabilities of observing a click on detectors when the inputs are equal and different, which affects the selection of the thresholds. The detailed analysis is in the Appendix \ref{C} and the results are shown in Table \ref{t18}.

	\begin{table}
		\centering
		\caption{Determine the relationship among four inputs by four-party quantum fingerprinting protocol of asymmetric channels. The $\alpha_k$ of each sender and the thresholds $C_2^{\rm{th}}$, $C_3^{\rm{th}}$, $C_4^{\rm{th}}$ can be different in each run and we write them in the corresponding 1st, 2nd, 3rd rows of the following table, respectively. $n=10^{14}$, $c=0.2$, $P_d=10^{-11}$, $\delta=0.22$, $\sqrt{\eta_1}=0.3$, $\sqrt{\eta_2}=0.4$, $\sqrt{\eta_3}=0.5$, $\sqrt{\eta_4}=0.6$, $\varepsilon= 10^{-5}$. $C_{\rm{o}}^{\rm{AE}}=1.01\times 10^{11}$, $C_{\rm{l}}^{\rm{AE}}=1.19\times 10^{7}$ .}
		\begin{tabular}{p{0.7cm}p{0.7cm}p{0.7cm}p{0.7cm}p{0.7cm}p{0.8cm}p{0.8cm}p{0.8cm}p{1.4cm}}
			\hline 
			&$\alpha_1$& $ \alpha_2 $ &$\alpha_3$& $ \alpha_4 $ & $C_{2}^{\rm{th}}$& $C_{3}^{\rm{th}}$  & $C_{4}^{\rm{th}}$&$Q^{\rm{R}}$ \\
			\hline 
			\hline                 
			1st&$109$& $109$ &$69$& $69$ & $5367$& $5700$ & $5332$&$4.43\times 10^6$\\                 
			2nd&$97$& $77$ &$99$& $78$ & $5519$& $5600 $ & $5439$\\       
			3rd&$90$& $84$ &$85$& $91$ & $5699$& $5600$ &$5347$ \\
			\hline
		\end{tabular}\label{t18}
		
	\end{table}

In the process of parameter optimization, it should be noted that the fingerprint states and thresholds of each run can be different, as long as $P_e\le \varepsilon$. As shown in table \ref{t18}, when $\sqrt{\eta_1}=0.3$, $\sqrt{\eta_2}=0.4$, $\sqrt{\eta_3}=0.5$, $\sqrt{\eta_4}=0.6$, on the first run of the protocol, the total amplitudes and thresholds can be set to $\alpha_1=109$, $\alpha_2=109$, $\alpha_3=69$, $\alpha_4=69$, $C_2^{\rm{th}}=5367$, $C_3^{\rm{th}}=5700$, $C_4^{\rm{th}}=5332$. If the relationship cannot be determined by $\rm{R1}$, then the interference positions of $S_2$ and $S_3$ needs to be exchanged and we need to run the protocol again. On the second run of the protocol, the total amplitudes and thresholds can be set to $\alpha_1=97$, $\alpha_2=77$, $\alpha_3=99$, $\alpha_4=78$, $C_2^{\rm{th}}=5519$, $C_3^{\rm{th}}=5600$, $C_4^{\rm{th}}=5439$. If $f^{\rm{R}}$ still cannot be calculated based on $\rm{R1}$ and $\rm{R2}$, then positions of $S_3$ and $S_4$ need to be exchanged. On the third run of the protocol, the total amplitudes and thresholds can be set to $\alpha_1=90$, $\alpha_2=84$, $\alpha_3=85$, $\alpha_4=91$, $C_2^{\rm{th}}=5699$, $C_3^{\rm{th}}=5600$, $C_4^{\rm{th}}=5347$.
We can find that the maximal total communication complexity $Q^{\rm{R}}=4.43\times 10^6$  is five orders of magnitude less than $C_{\rm{o}}^{\rm{AE}}=1.01\times 10^{11}$ of classical optimal fingerprinting. Moreover, it breaks the classical limit $C_{\rm{l}}^{\rm{AE}}=1.19\times 10^7$. In short, they satisfy $Q^{\rm{R}}<C_{\rm{l}}^{\rm{AE}}<C_{\rm{o}}^{\rm{AE}}$ and combined with \eqref{QC}, it can be concluded that $Q^{\rm{R}}<C_{\rm{l}}^{\rm{R}}<C_{\rm{o}}^{\rm{R}}$. This shows the advantage that the quantum fingerprint network can reduce the communication complexity compared with the classical version.

	
	\section{Discussion}
	In this work, we demonstrate the advantages of quantum fingerprinting networks in the field of communication complexity compared with classical versions according to $Q^{\rm{R}}<C_{\rm{l}}^{\rm{AE}}<C_{\rm{o}}^{\rm{AE}}$.  It is important to note that the actual advantages (the difference among $Q^{\rm{R}}$, $C_{\rm{l}}^{\rm{R}}$ and $C_{\rm{o}}^{\rm{R}}$) will be greater than that shown by the above comparison method because $C_{\rm{o}}^{\rm{R}}>C_{\rm{o}}^{\rm{AE}}$ and $C_{\rm{l}}^{\rm{R}}>C_{\rm{l}}^{\rm{AE}}$.  
	
    \textbf{The comparison between multi-party protocols and two-party protocols.}
    For simplicity, we call the multi-party quantum fingerprinting protocol the multi-party protocol and we call the protocol, which uses the two-party quantum fingerprinting to compare $N$ inputs in pairs, the two-party protocol.

    If the Referee needs to calculate $f^{\rm{R}}$, when the relationship is AABB (or AAAA), the four-party protocol only needs to be run once, $t_{\rm{M}}=1$, $Q^{\rm{R}}=1.55\times 10^6 $ (the parameters are shown in Table \ref{t18} and $Q^{\rm{R}}$ is the communication complexity of the first run of the four-party protocol), while the two-party protocol needs to run multiple times, $t_{\rm{T}}=4$ (or 3), $Q^{\rm{R}}=3.19\times 10^6$ (or $2.75\times 10^6$). Therefore, the four-party protocol shows advantages in terms of communication time and communication complexity.
    As shown in Table \ref{t12} in Appendix \ref{D2}, it can be found that no matter what the relationship is, the number of runs $t_{\rm{M}}$ required by the four-party protocol is lower than $t_{\rm{T}}$ required by the two-party protocol.

    \begin{table}
    	\centering
    	\caption{The comparison of the $N$-party quantum fingerprinting and two-party protocol when $N=2^s, s\in \mathbb{Z^+}$. $QF_N^{\rm{M}}$ refers to the $N$-party quantum fingerprinting protocol based on balanced BSs and $QF_N^{\rm{T}}$ refers to the two-party protocol, which compares $N$ inputs in pairs. $f$ is the function that the protocol needs to compute and $t_{\rm{max}}$ represents the maximum number of times the protocol needs to be run.}
    	\begin{tabular}{p{1.8cm}p{1.8cm}p{3cm}}
    		\hline 
    		& $ f $ & $ t_{\rm{max}} $   \\
    		\hline 
    		\hline                 
    		$QF_N^{\rm{M}}$& $f^{\rm{AE}}$ & $1$ \\
    		$QF_N^{\rm{T}}$& $f^{\rm{AE}}$ & $N-1$  \\
    		$QF_N^{\rm{M}}$& $f^{\rm{R}}$  & $N-1$ \\
    		$QF_N^{\rm{T}}$& $f^{\rm{R}}$ & $\frac{1}{2}N(N-1)$ \\
    		\hline 
    	\end{tabular}\label{t20}
    \end{table}

    Furthermore, in Table \ref{t20}, we compare the $N$-party quantum fingerprinting  based on balanced BSs and the protocol based on two-party fingerprinting when $N=2^s,s\in \mathbb{Z^+}$. When calculating $f^{\rm{AE}}$, the multi-party protocol only needs to run once, while the two-party scheme needs to be run up to $(N-1)$ times. When calculating $f^{\rm{R}}$, the multi-party protocol only needs to be run up to $(N-1)$ times, while the two-party scheme needs to run up to $\frac{1}{2}N(N-1)$ times. To sum up, the multi-party protocol has obvious advantages over the two-party protocol, especially in terms of communication time.
    
    \textbf{Multi-bits encoding Methods.}
    To further improve the performance of the protocol, multiple bits of $E(x_k)$ can be encoded on each coherent state \cite{lovitz2018families}.
    Different from \eqref{e1}, the fingerprint state of the $k$th sender is prepared in the following form
    
    \begin{equation}\label{}
    \begin{split}
    &\ket{s_k}=\bigotimes_{j=1,\rm{odd}}^{m}\biggl | \mathrm{i}^{E(x_k)_j\oplus E(x_k)_{j+1}}(-1)^{E(x_k)_j}\frac{\alpha_k}{\sqrt{m/2}}\biggr>_j,\\
    \end{split}
    \end{equation}
    where  $\rm{i}^2=-1$ and $E(x_k)_j\oplus E(x_k)_{j+1}$ means $E(x_k)_j+ E(x_k)_{j+1}$ $\mod 2$.
    
    \begin{table}
    	\centering
    	\caption{The selection of parameters in two-party quantum fingerprinting protocol of asymmetric channels when using the two-bits encoding method. $n=3\times10^{12}$, $c=0.2$, $P_d=10^{-10}$, $\delta=0.22$, $\sqrt{\eta_1}=0.3$, $\sqrt{\eta_2}=0.4$, $\varepsilon= 10^{-5}$.}
    	\begin{tabular}{p{1.5cm}p{1.5cm}p{1.5cm}p{2cm}}
    		\hline 
    		 $\alpha_1$ & $ \alpha_2 $ & $ C_{2}^{\rm{th}} $ & $ Q^{\rm{R}} $ \\
    		\hline 
    		\hline                 
    		$69$& $70$ & $898$& $ 3.91\times 10^5$\\
    		\hline
    	\end{tabular}\label{t131}
    \end{table}

The total communication complexity of the single-bit encoding method is $Q^{\rm{R}}_{\rm{single-bit}}=5.52\times 10^5$ (as shown in the table \ref{t15}), while that of the two-bit encoding method is smaller, $Q^{\rm{R}}_{\rm{two-bits}}=3.91\times 10^5$, as shown in table \ref{t131}. In either case, the total communication complexity of the quantum fingerprinting is about five orders of magnitude smaller than that of the classical optimal protocol, $C_{\rm{o}}^{\rm{AE}}=1.24\times 10^{10}$. Moreover, the communication complexity of both schemes breaks the classical limit $C_{\rm{l}}^{\rm{AE}}=1.46\times 10^6$. They satisfy $Q^{\rm{R}}_{\rm{two-bits}}<Q^{\rm{R}}_{\rm{single-bit}}<C_{\rm{l}}^{\rm{AE}}<C_{\rm{o}}^{\rm{AE}}$, combined with \eqref{QC}, we can conclude that $Q^{\rm{R}}_{\rm{two-bits}}<Q^{\rm{R}}_{\rm{single-bit}}<C_{\rm{l}}^{\rm{R}}<C_{\rm{o}}^{\rm{R}}$, which reflects the advantage of quantum fingerprinting in communication complexity. Besides, in the two-bits encoding method, the number of fingerprinting states sent by the senders is half that of the single-bit encoding method, which means in terms of communication time, the two-bit encoding method also has more advantages.

Of course, it does not mean that the more bits are encoded in each coherent state, the better. If all bits are encoded in one coherent state, the communication time is undoubtedly minimal, but the error probability $P_e$ may be very high. Therefore, it is important to find a balance among the communication complexity, error probability and communication time. 
   
    
   Our analysis above is based on ideal balanced BSs. In practice, the interference visibility $\nu<1$ ($\nu=0.99$ in \cite{xu2015experimental}), but this does not affect the analysis of decision rules. We consider the impact of interference visibility $\nu$ in the Appendix D. It can be found that the there is a slight effect on parameter selections and the communication complexity $Q^{\rm{R}}$ increase a little, but they still satisfy $Q^{\rm{R}}<C_{\rm{l}}^{\rm{R}}<C_{\rm{o}}^{\rm{R}}$.
   This means that the quantum fingerprinting still has obvious advantages over the classical version in terms of communication complexity after considering the effect of $\nu$.
    

    The device composed of balanced BS used in $N$-party quantum fingerprinting, $N=2^s,s\in \mathbb{Z^+}$, can also be used in $M$-party quantum fingerprinting, $M<N$, which is analyzed in the Appendix E. Compared with the extendable design for $M$-party quantum fingerprinting ($M=3$) \cite{GSMultiparty2020}, we only need balanced BSs and the number of runs needed to compute $f^{\rm{R}}$ is smaller, which is very beneficial to the experiment.
    
    When the Referee needs to calculate $f^{\rm{AE}}$, the protocol only needs to be run once. In addition to the two decision rules proposed in \cite{GSMultiparty2020}, we provide another decision rule to calculate $f^{\rm{AE}}$. As shown in the second column $\rm{R1}$ in Table \ref{t9}, the Referee can only observe two detectors $\rm{D2}, \rm{D3}$ (or $\rm{D3}, \rm{D4}$) and only if $C_2$, $C_3$ (or $C_3$, $C_4$) are all less than the corresponding threshold respectively, $f^{\rm{AE}}=1$; otherwise, $f^{\rm{AE}}=0$.
    
    The characteristic of our method of exchanging interference positions to determine the relationship among $N$ inputs is that the latter run depends on the results of the previous runs. When $N$ is larger, we may be able to design more flexible steps. For example, according to the results of the previous runs, the latter run only requires some of the $N$ senders to send fingerprint states for comparison, so as to further reduce the communication complexity.


 \section{Conclusion}

We provide a general theory of quantum fingerprinting network, which can determine the relationship among multiple messages, and choose the optimal parameters to minimize the communication complexity in the case of asymmetric channels. We take the four-party quantum fingerprinting protocol as an example to analyze in detail how to calculate $f^{\mathrm{R}}$ by exchanging the interference position of different senders. Moreover, we demonstrate the advantages of quantum networks, especially in terms of communication time, by comparing quantum fingerprinting networks with the two-party protocol. To further improve the performance of the protocol, we use the multi-bits scheme, in which multiple bits are encoded in each coherent state. It is important to find a good balance among communication complexity, error probability and communication time.

				
\appendix

\section{The expression for $k_{t}^{i,j}$ } \label{A}

There are $j$ groups and $G_j$ represents the same $N_j$ inputs of $N$ inputs which satisfies $N_1\ge N_2\ge\dots\ge N_j$, $j=1,2,\dots,N$. We define $N_1=i$, which means at most $i$ inputs are the same, $i=N-(j-1),N-(j-1)-1,\dots,\lceil \frac{N}{j} \rceil$ . $k=1,2,\dots,k_{t}^{i,j}$, where $k_{t}^{i,j}$ is the total number of cases for the same $(i,j)$,

\begin{equation}\label{}
\begin{split}
k_{t}^{i,j}=\sum_{i\ge N_2\ge N_3\ge \dots \ge N_j}\frac{1}{s_{\mathbb{G}}}C_N^iC_{N-i}^{N_2}C_{N-i-N_2}^{N_3}\dots C_{N_{j-1}+N_j}^{N_{j-1}},\\
\end{split}
\end{equation}
where $N_2+N_3+\dots+N_j=N-i$; $s_{\mathbb{G}}$ represents the effect of the repeat count due to the same number of elements in different groups. For example, when $N=4$, $i=2$, $j=2$, $(x_1,x_2)_1,(x_3,x_4)_2$ and $(x_3,x_4)_1,(x_1,x_2)_2$ represent the same relationship. So, we need to use $s_{\mathbb{G}}$ to eliminate the effect of this type of repeat count.

When $N$ is large, the relationship among $N$ inputs is complicated, i.e., $f^{\rm{R}}$ has many values.
For instance, when $N=8$, $i=4$, $j=3$, $k_{t}^{3,4}=C_8^4C_4^3+\frac{1}{2}C_8^4C_4^2=490.$

\section{ Quantum fingerprinting protocol of asymmetric channels} \label{C}

 Whether the channel is symmetrical does not change the decision rules of the Referee. However, it affects the expression of the probability of observing a click on the detectors for each coherent state when the inputs are equal and different. Therefore, it has an impact on choosing the appropriate fingerprint states and thresholds to minimize communication complexity. So, we focus on analyzing the probability of observing a click on the detectors for each coherent state in different protocols.
 
\begin{center}
\textbf{Two-party quantum fingerprinting protocol of asymmetric channels}
\end{center}

\setcounter{figure}{0} 
\renewcommand{\thefigure}{C\arabic{figure}}

 When $N=2$, there are two detectors $\rm{D_1}$ and $\rm{D_2}$ on the Referee and ideally, when $x_1=x_2$, only $\rm{D_1}$ has clicks; when $x_1\neq x_2$, both $\rm{D_1}$ and $\rm{D_2}$ have clicks. The probabilities of observing a click for each pulse sent on $\rm{D_2}$ are $P_2^{\rm{E}}$ and $P_2^{\rm{D}}$ when the input information is equal and different.

\begin{equation}\label{C1}
\begin{split}
P_{2,\rm{single}}^{\rm{E}}&=\big[1-e^{-\frac{1}{2m}(\sqrt{\eta_1}\alpha_1-\sqrt{\eta_2}\alpha_2)^2}\big]+P_d\\
P_{2,\rm{single}}^{\rm{D}}&=\delta\big[1-e^{-\frac{1}{2m}(\sqrt{\eta_1}\alpha_1+\sqrt{\eta_2}\alpha_2)^2}\big]+\\
&\left(1-\delta\right)\big[1-e^{-\frac{1}{2m}(\sqrt{\eta_1}\alpha_1-\sqrt{\eta_2}\alpha_2)^2}\big]+P_d.\\
\end{split}
\end{equation}	

In addition, it can also be determined by the detector $D_1$. The analysis is similar.

To further improve the performance of the protocol, we consider the multi-bits encoding scheme. The following is the corresponding probability of two-bits encoding scheme

\begin{equation}\label{}
\begin{split}
P_{2,\rm{two}}^{\rm{E}}&=\big[1-e^{-\frac{1}{m}(\sqrt{\eta_1}\alpha_1-\sqrt{\eta_2}\alpha_2)^2}\big]+P_d\\
P_{2,\rm{two}}^{\rm{D}}&=\left(1-\delta\right)^2\big[1-e^{-\frac{1}{m}(\sqrt{\eta_1}\alpha_1-\sqrt{\eta_2}\alpha_2)^2}\big]+\\
&2\delta\left(1-\delta\right)\big[1-e^{-\frac{1}{m}\left((\sqrt{\eta_1}\alpha_1)^2+\left(\sqrt{\eta_2}\alpha_2\right)^2\right)}\big]+\\
&\delta^2\big[1-e^{-\frac{1}{m}\left(\sqrt{\eta_1}\alpha_1+\sqrt{\eta_2}\alpha_2\right)^2}\big]+P_d.\\
\end{split}
\end{equation}

\begin{center}
	\textbf{Four-party quantum fingerprinting protocol of asymmetric channels}
\end{center}

Similar to the decision rules shown in table \ref{t9}, we can determine the relationship among four inputs according to the detector $\rm{D_2}$, $\rm{D_3}$ and $\rm{D_4}$. The probability of observing a click on $\rm{D_2}$, $\rm{D_3}$ and $\rm{D_4}$ for each pulse when the inputs are equal and different are as follows

\begin{equation}\label{}
\begin{split}
P_{2}^{\rm{E},s}&=\big[1-e^{-\frac{1}{2m}\left(-\sqrt{\eta_i}\alpha_i^s+\sqrt{\eta_j}\alpha_j^s\right)^2}\big]+P_d\\
P_{2}^{\rm{D},s}&=\delta\big[1-e^{-\frac{1}{2m}\left(\sqrt{\eta_i}\alpha_i^s+\sqrt{\eta_j}\alpha_j^s\right)^2}\big]+\\
&\left(1-\delta\right)\big[1-e^{-\frac{1}{2m}\left(-\sqrt{\eta_i}\alpha_i^s+\sqrt{\eta_j}\alpha_j^s\right)^2}\big]+P_d\\
P_{3}^{\rm{E},s}&=\big[1-e^{-\frac{1}{4m}\left(-\sqrt{\eta_i}\alpha_i^s-\sqrt{\eta_j}\alpha_j^s+\sqrt{\eta_k}\alpha_k^s+\sqrt{\eta_l}\alpha_l^s\right)^2}\big]+P_d\\
P_{3}^{\rm{D},s}&=\delta\big[1-e^{-\frac{1}{4m}{(x^s)^2}}\big]+\\
&\left(1-\delta\right)\big[1-e^{-\frac{1}{4m}(-\sqrt{\eta_i}\alpha_i^s-\sqrt{\eta_j}\alpha_j^s+\sqrt{\eta_k}\alpha_k^s+\sqrt{\eta_l}\alpha_l^s)^2}\big]+P_d\\
P_{4}^{\rm{E},s}&=\big[1-e^{-\frac{1}{2m}\left(-\sqrt{\eta_k}\alpha_k^s+\sqrt{\eta_l}\alpha_l^s\right)^2}\big]+P_d\\
P_{4}^{\rm{D},s}&=\delta\big[1-e^{-\frac{1}{2m}\left(\sqrt{\eta_k}\alpha_k^s+\sqrt{\eta_l}\alpha_l^s\right)^2}\big]+\\
&\left(1-\delta\right)\big[1-e^{-\frac{1}{2m}(-\sqrt{\eta_k}\alpha_k^s+\sqrt{\eta_l}\alpha_l^s)^2}\big]+P_d,\\
\end{split}
\end{equation}
where $s=1,2,3$,  $\left(i,j\right)\left(k,l\right)$$=\left(1,2\right)\left(3,4\right)$, $\left(1,3\right)$$\left(2,4\right)$,
$\left(1,4\right)$$\left(2,3\right)$ and $x^s=\min \{|\sqrt{\eta_i}\alpha_i^s-\sqrt{\eta_j}\alpha_j^s+\sqrt{\eta_k}\alpha_k^s+\sqrt{\eta_l}\alpha_l^s|, |-\sqrt{\eta_i}\alpha_i^s+\sqrt{\eta_j}\alpha_j^s+\sqrt{\eta_k}\alpha_k^s+\sqrt{\eta_l}\alpha_l^s|, |-\sqrt{\eta_i}\alpha_i^s-\sqrt{\eta_j}\alpha_j^s-\sqrt{\eta_k}\alpha_k^s+\sqrt{\eta_l}\alpha_l^s|, |-\sqrt{\eta_i}\alpha_i^s-\sqrt{\eta_j}\alpha_j^s+\sqrt{\eta_k}\alpha_k^s-\sqrt{\eta_l}\alpha_l^s| \}$. This means that we can change the amplitude of coherent states and thresholds  of detectors in each run with the purpose of minimizing $Q^{\rm{R}}$ under the condition that $P_e\le\varepsilon$.

\section{ Compare the number of runs between a multi-party protocol and a two-party protocol } \label{D2}

\setcounter{table}{0} 
\renewcommand{\thetable}{C\arabic{table}}

We can use a two-party quantum fingerprinting protocol to perform pairwise comparison to determine the relationship among four inputs, as shown in Table \ref{t12}. The two-party protocol must be run at least three times and at most six times to determine $f^{\rm{R}}$. The blank sections in the table indicate that this run is not required. The number $0$ means the total counts are less than the threshold, and the number $1$ means the total counts are greater than the threshold. $t_{\rm{T}}$ stands for the number of times to be run when using the two-party protocol, $t_{\rm{M}}$ stands for the number of times to be run when using the four-party quantum fingerprinting protocol of which the decision rules are based on three detectors $\rm{D_2}$, $\rm{D_3}$ and $\rm{D_4}$. 

It can be seen from the Table \ref{t12} that no matter what the relationship is, the number of runs $t_{\rm{M}}$ required by the four-party protocol is lower than $t_{\rm{T}}$ required by the two-party protocol. For example, if the relationship is AABB, the protocol needs to be run four times, $t_{\rm{T}}=4$, using two-party protocol. However, the protocol only needs to be run once, $t_{\rm{M}}=1$, to identify AABB by four-party protocol.

\begin{table}
	\centering
	\caption{Compare the number of runs $t_{\rm{M}}$ using the four-party protocol of which the decision rules are based on three detectors and $t_{\rm{T}}$ using the two-party protocol. $\rm{R1}$, $\rm{R2}$, $\rm{R3}$, $\rm{R4}$, $\rm{R5}$ and $\rm{R6}$ respectively represents the comparison results of $E(x_1)$ and $E(x_2)$, $E(x_1)$ and $E(x_3)$, $E(x_1)$ and $E(x_4)$, $E(x_2)$ and $E(x_3)$, $E(x_2)$ and $E(x_4)$, $E(x_3)$ and $E(x_4)$ with two-party protocol. The number $0$ means the total counts are less than the threshold, and the number $1$ means the total counts are greater than the threshold. The blank sections indicate that this run is not required.}
	\begin{tabular}{p{1.4cm}p{0.7cm}p{0.7cm}p{0.7cm}p{0.7cm}p{0.7cm}p{0.7cm}p{0.7cm}p{0.7cm}}
		\hline
		& $ \rm{R1}$ & $ \rm{R2} $ & $ \rm{R3}$ & $ \rm{R4}$ & $ \rm{R5} $ & $ \rm{R6}$& $t_{\rm{T}}$& $t_{\rm{M}}$ \\
		\hline
		\hline                  
		AAAA & $0$ & $0$ &$0$ & & & &$3$ &$1$\\
		AAAB & $0$ & $0$ &$1$ & & & &$3$ &$2$\\
		AABA & $0$ & $1$ & $0$& & & &$3$ &$2$\\
		ABAA & $1$ & $0$ & $0$& & & &$3$ &$2$\\
		BAAA & $1$ & $1$ & $1$& $0$ & $0$ &&$5$ &$2$ \\
		AABB & $0$ & $1$ & $1$& &&$0$ &$4$ &$1$\\
		ABAB & $1$ & $0$ & $1$& & $0$& &$4$ &$2$ \\
		ABBA & $1$ & $1$ & $0$& $0$ && &$4$ &$2$\\
		AABC & $0$ & $1$ & $1$& &&$1$&$4$ &$2$ \\
		ABAC & $1$ & $0$ & $1$& &$1$ &&$4$ &$2$ \\
		ABCA & $1$ & $1$ & $0$ & $1$ & &&$4$ &$3$\\
		BAAC & $1$ & $1$ & $1$ & $0$ & $1$&&$5$ &$3$\\
		BACA & $1$ & $1$ & $1$ & $1$ &$0$ &&$5$ &$2$\\
		BCAA & $1$ & $1$ & $1$ & $1$ &$1$ &$0$&$6$ &$2$\\		
		ABCD & $1$ & $1$ & $1$ & $1$ &$1$ &$1$&$6$ &$2\ \rm{or} \ 3$\\	
		\hline          
	\end{tabular}\label{t12}
\end{table}

\section{ The impact of interference visibility} \label{D}

The protocols we analyzed are based on ideal BS with interference visibility $\nu=1$. In practice, the interference visibility is slightly lower, for example $\nu=0.99$ in \cite{xu2015experimental}. We can make some modifications to the probability of observing a click on detectors when considering the impact of $\nu$. Take the two-party quantum fingerprinting as example. 

The probability of observing a click on detectors $\rm{D_2}$ for each pulse when the inputs are equal and different becomes 
		
\begin{equation}\label{D1}
\begin{split}
P_{2}^E&=\nu\big[1-e^{-\frac{1}{2m}\left(\sqrt{\eta_1}\alpha_1-\sqrt{\eta_2}\alpha_2\right)^2}\big]+\\
&\left(1-\nu\right)\big[1-e^{-\frac{1}{2m}(\sqrt{\eta_1}\alpha_1+\sqrt{\eta_2}\alpha_2)^2}\big]+P_d,\\
\end{split}
\end{equation}

\begin{equation}\label{}
\begin{split}
P_{2}^D&=\delta\big[\nu\left(1-e^{-\frac{1}{2m}(\sqrt{\eta_1}\alpha_1+\sqrt{\eta_2}\alpha_2)^2}\right)+\\
&\left(1-\nu\right)\left(1-e^{-\frac{1}{2m}(\sqrt{\eta_1}\alpha_1-\sqrt{\eta_2}\alpha_2)^2}\right)\big]+\\
&\left(1-\delta\right)\big[\nu\left(1-e^{-\frac{1}{2m}\left(\sqrt{\eta_1}\alpha_1-\sqrt{\eta_2}\alpha_2\right)^2}\right)+\\
&\left(1-\nu\right)\left(1-e^{-\frac{1}{2m}(\sqrt{\eta_1}\alpha_1+\sqrt{\eta_2}\alpha_2)^2}\right)\big]+P_d.\\
\end{split}
\end{equation}

\setcounter{table}{0} 
\renewcommand{\thetable}{D\arabic{table}}

\begin{table}
	\centering
	\caption{The selection of parameters in two-party quantum fingerprinting protocol with non-ideal BS in the case of asymmetric channels. $n=3\times10^{12}$, $c=0.2$, $P_d=10^{-10}$, $\delta=0.22$, $\sqrt{\eta_1}=0.3$, $\sqrt{\eta_2}=0.4$, $\varepsilon=10^{-5}$. $C_{\rm{o}}^{\rm{AE}}=1.24\times 10^{10}$, $C_{\rm{l}}^{\rm{AE}}=1.46\times 10^{6}$.}
	\begin{tabular}{p{1.5cm}p{1.5cm}p{1.5cm}p{1.5cm}p{1.5cm}}
		\hline 
		$\nu$&$\alpha_1$& $ \alpha_2 $ & $ C_{2}^{\rm{th}} $ & $ Q^{\rm{R}} $ \\
		\hline 
		\hline                 
		$0.99$ & $88$& $77$ & $1695$ & $5.67\times 10^5$ \\
		\hline
	\end{tabular}\label{t21}
\end{table}

The comparison of the first row in Table \ref{t15} and Table \ref{t21} shows that there is a slight effect on parameter selections after considering the interference visibility $\nu$ and the communication complexity $Q^{\rm{R}}$ increases a little, but they still satisfy $Q^{\rm{R}}<C^{\rm{AE}}_{\rm{l}}<C^{\rm{AE}}_{\rm{o}}$. Based on this, and combined with \eqref{QC}, we can obtain $Q^{\rm{R}}<C_{\rm{l}}^{\rm{R}}<C_{\rm{o}}^{\rm{R}}$, which means that the communication complexity of the quantum fingerprinting protocol is lower than that of the classical version. 

The analysis in the multi-party quantum fingerprinting protocol is similar.

\section{$M$-party quantum fingerprinting based on the device  composed of balanced BSs} \label{E}

The device  composed of balanced BSs, which can be used for $N$-party quantum fingerprinting, $N=2^s,s\in \mathbb{Z^+}$, can also be used for $M$-party quantum fingerprinting, where $M<N$. 
We show how to determine the relationship among three inputs using the device shown in Fig.\ref{f1}, which is the device used for four-party quantum fingerprinting.

Different from four-party quantum fingerprinting, when $N=3$, the sender $S_1$ prepares the same two sets of fingerprint states according to $E(x_1)$, one is still sent to the interference position of $S_1$ and the other is sent to the interference position of $S_4$ in the Fig.\ref{f1}. The Referee can determine the relationship among the three inputs according to the comparison results of the total counts $C_2$, $C_3$, $C_4$ and corresponding thresholds $C_2^{\rm{th}}$, $C_3^{\rm{th}}$, $C_4^{\rm{th}}$, respectively. As shown in Table \ref{t13}, each comparison result corresponds to a unique relationship, so the Referee only needs to run the protocol once to calculate $f^{\rm{R}}$ . For example, if the results is R1=$011$, then the relationship is AAB.

The three-party quantum fingerprinting based on extendable design \cite{GSMultiparty2020} needs to exchange the interference position to determine the relationship, which can only be completed for three times at most. The method above can calculate $f^R$ in one run and does not require unbalanced BSs, which is very beneficial for experiments.

\setcounter{table}{0} 
\renewcommand{\thetable}{E\arabic{table}}
\begin{table}
	\centering
	\caption{Determine the relationship among the three inputs $x_1,x_2,x_3$ by detectors $\rm{D_2}$, $\rm{D_3}$ and $\rm{D_4}$. $\rm{R1}$ represents the comparison results of total counts $C_2$, $C_3$, $C_4$ and thresholds $C_{2}^{\rm{th}}$, $C_{3}^{\rm{th}}$, $C_{4}^{\rm{th}}$ of the three detectors after the first run of the protocol, respectively.  }
	\begin{tabular}{p{2cm}p{2cm}p{3.5cm}}
		\hline
		$x_1$,$x_2$,$x_3$,$x_1$ & $ R1$ & $f^{\rm{R}}$ \\
		\hline
		\hline                  
		AAAA & $000$ & $4:(x_1,x_2,x_3)_1$\\
		AABA & $011$ & $ 3:(x_1,x_2)_1,(x_3)_2$\\
		ABAA & $110$ & $ 2:(x_1,x_3)_1,(x_2)_2$\\	
		BAAB & $101$ & $ 1:(x_2,x_3)_1,(x_1)_2$\\
		ABCA & $111$ & $ 0:(x_1)_1,(x_2)_2,(x_3)_3$\\		
		\hline          
	\end{tabular}\label{t13}
\end{table}

\bibliography{refs}

\end{document}